\documentclass[eqsecnum,aps,epsfig]{revtex4}
\usepackage{amsmath,amssymb,amsthm,bm}
\usepackage{psfrag}
\usepackage[dvips]{graphicx}

\begin{document}
\title{ Role of the SU($2$) and SU($3$) subgroups in observing confinement in the G($2$) gauge group}
\author{S.~M.~Hosseini Nejad$^1$ and S.~Deldar$^2$}
\affiliation{
Department of Physics, University of Tehran, P.O. Box 14395/547, Tehran 1439955961,
Iran \\
$^1$smhosseininejad@ut.ac.ir\\
 $^2$sdeldar@ut.ac.ir
 }

\begin{abstract}

Applying the domain model to the G($2$) gauge group and the thick center vortex model to the SU($2$) and SU($3$) subgroups of G($2$), we calculate the 
potentials between static sources, as well as some parameters of the vortex profile. Comparing the results obtained from G($2$) and 
its subgroups, we argue that SU($2$) and SU($3$) gauge groups have important roles in observing confinement in the G($2$) gauge group.
\end{abstract}

\maketitle

\section{INTRODUCTION}
The mechanism of color confinement is still one of the unsolved and challenging problems in particle physics. It is hard because it is a nonperturbative
phenomenon and must be solved by nonperturbative techniques. Lattice gauge theory, as a numerical method, has been successful in predicting  
potentials between color sources and observing the confinement  in various representations \cite{Bali2000,Deld1999,Berna,Ambjorn,Michael,Poulis,Ohta,Greensite2007,Olejnik2008}. In addition, a variety of phenomenological models have been 
proposed to explain the confinement mechanism. According to these models, the vacuum of QCD is filled by some special class of field configurations
such as center vortices, monopoles, instantons and, etc. \cite{J. Green2010}. In this article, we focus on the thick center vortex model. Center vortices were introduced in the late $1970$'s
by 't Hooft \cite{Hooft1979}. Color confinement has been explained based on the condensation of center vortices in the vacuum of QCD. The center 
vortices are 
linelike (surfacelike) objects
in three (four) dimensions which carry a quantized magnetic flux in terms of the nontrivial center elements of the gauge group. The interactions between Wilson loops and center 
vortices in the fundamental representation lead to a linear potential but it is not possible to reproduce the
intermediate linear potential for the color sources in higher representations predicted by lattice 
calculations \cite{Bali2000,Deld1999}. The vortices were thickened in the thick center vortex model
 \cite{Fabe1998} and the linear potentials for the higher representations, as well as the fundamental representation have been observed. Later, in order to increase the
length of the linear part of the potential, Greensite $\it{et~ al.}$ \cite{Greensite2007} have  modified the model by using both the trivial and nontrivial
center elements. Studying the modified thick center vortex model in a group, using only the trivial center element seems to be very interesting. 
This is because as mentioned above, the magnetic fluxes carried by the center vortices are quantized in terms of the nontrivial center elements of the group. 
Therefore, one does not expect confinement in a group which does not contain any nontrivial center element. In other words, no center element 
means no confinement.
However this is in contrast with the lattice results of the G($2$) gauge group \cite{Olejnik2008}, as an example of a group without any nontrivial center element. A linear potential 
for the intermediate distances has been observed in lattice calculations. G($2$) is the simplest exceptional Lie group which has only a trivial center element and 
its universal covering group is itself.
In language of homotopy group, 
the first  homotopy group shows that center vortices are absent in the theory, i.e., 
\begin{equation}
\pi_1(G(2)/\mathbb I) = \mathbb I.
\end{equation}
Thus, G($2$) gauge theory is a good laboratory for
studying the role of the trivial center element for color confinement.

Recently, G($2$) Yang Mills theory has attracted considerable attention for the confinement problem in QCD  \cite{Greensite2007, Olejnik2008, Holland2003, Pepe2006, Pepew2007, Cossu2006, Cossum2007, Cossud2007, Maas2008}.
In our previous article \cite{Deldar2012}, we have calculated the potentials between two G($2$) heavy sources in the fundamental, adjoint and $27$ dimensional representations, by the thick 
center vortex model with the idea of the domain structure. In agreement with lattice results \cite{Olejnik2008}, we have observed screening of the sources for the large distances and linear potentials at intermediate distances roughly proportional 
with the Casimir ratios. Screening is expected, since the only center element of G($2$) is trivial and it does not have any contribution
to the Wilson loop at large distances where it is located completely inside the loop. We have discussed the possible reasons of the observed linear potential at intermediate
distances. We have argued that the thickness of the domains and the SU($3$) subgroup of the G($2$) gauge group may be responsible for this linear behavior. 

In this paper, we discuss the role of the nontrivial centers of the SU($2$) and SU($3$) subgroups of G($2$) in observing confinement in the G($2$) gauge group. In the next section, we give a brief 
summary on the thick center vortex model and the domain
structure model. Some general properties of the G($2$) gauge group are studied in Sec. III.
Then, in Secs. IV and V, we investigate the reasons of the confinement in G($2$). First, the domain model is applied to the G($2$) gauge group and then the thick center vortex 
model to the SU($2$) and SU($3$) subgroups of the G($2$) gauge group and then by comparing the results we discuss the reasons of
confinement in the G($2$). Next, we study ${\mathrm {Re}}(g_{r})$, a function of the vortex profile,  for the fundamental and adjoint representations of G($2$) and compare its extremums with the ones 
of the SU($2$) and SU($3$) subgroups of G($2$). By comparing the results of these two parts, we conclude that SU($2$) and SU($3$) subgroups of G($2$) have important roles in 
observing confinement in G($2$).

\section{Thick Center Vortex Model including trivial domains }

Postulating a kind of domain structure in the vacuum, the thick center vortex model has been modified and the length of the linear regime at intermediate distances has been increased \cite{Greensite2007}. This model, sometimes called the domain model, assumes that the vacuum of quantum chromodynamics is filled with two types of domains called center vortices and vacuum 
domains. Magnetic flux in each domain is quantized in terms of the center elements which are trivial for the vacuum domain and nontrivial for the vortices. Therefore, vacuum-type domains carry a zero total magnetic flux in contrast with the center vortex type domains. In SU($N$) gauge theory, there are $N-1$ types of center vortices and one type of vacuum domain. For example for the SU($2$) gauge group, with the 
center elements $I$ and $-I$, there are two types of domains: $I$ and $-I$. The first one corresponds to the vacuum domain and the second one corresponds to the nontrivial center vortex. The 
induced potential between static sources is \cite{Fabe1998,Greensite2007}
\begin{equation}
\label{potential}
V(R) = \sum_{x}\ln\left\{ 1 - \sum^{N-1}_{n=0} f_{n}
(1 - {\mathrm {Re}} g_{r} [\vec{\alpha}^n_{C}(x)])\right\}
\end{equation}
$x$ is the location of the center of the vortex and  $f_{n}$ is the
probability that any given unit area is pierced by a domain of type $n$. $n=0$ indicates the vacuum domain and $n=1,...,N-1$ represent the center vortices or the nontrivial domains. $g_r[\vec{\alpha}]$ gives the information about the flux distribution and
the contribution that a domain with its center in a specific plaquette may have to the Wilson loop. It is given by 
\begin{equation}
\label{gr1}
g_r[\vec{\alpha}^n(x)]=\frac{1}{d_r}\mathrm{Tr}(\exp[i\vec{\alpha}^n(x) \cdot \vec{H}])
\end{equation}
where $\{H_{i},i=1,2,...,N-1\}$  are the generators spanning the Cartan subalgebra, $d_r$ is the dimension of the
representation $r$ and $\vec{\alpha}^n(x)$ shows the flux profile for the domain of type $n$. 
If the domain is completely contained within the Wilson loop area, then
\begin{equation}
\label{alfa1}
\exp(i\vec{\alpha}^{(n)}\cdot\vec{H})=z_{n}I
\end{equation}
where
\begin{equation}
z_n=e^{\frac{2\pi in}{N}} \in \mathbb Z_N
\end{equation}
and $I$ is the unit matrix.
 The normalization constant  $\vec{\alpha}^n(x)$ is obtained from the above maximum flux condition. For the G($2$) gauge group, since the center group contains only the trivial element, all domains are of the vacuum type. We apply the domain model to the G($2$) gauge group and the thick center vortex model to the SU($2$) and SU($3$) subgroups of G($2$) and then by comparing the results we discuss the reasons of observing 
confinement in the G($2$) gauge group. In the next section, we present some general properties of the G($2$) gauge group.

\section{G($2$) Gauge Group}
The G($2$) exceptional Lie group may be constructed as a subgroup of the real group SO($7$) which has twenty-one $7 \times 7$ real orthogonal generators. The rank of G($2$) is $2$ and the rank of SO($7$) is $3$. In addition to the usual properties of $SO(7)$ matrices
\begin{equation}
  \det U = 1 \qquad U^{-1} = U^{T},
\end{equation}
the G($2$) group elements  satisfy another constraint:
\begin{equation}
T_{abc}=T_{def}U_{da}U_{eb}U_{fc}
\label{constraint}
\end{equation}
where $T$ is a total antisymmetric tensor and its nonzero elements are \cite{Cacciatori:2005yb}
\begin{equation}
T_{127} = T_{154} = T_{163} = T_{235} = T_{264} = T_{374} = T_{576} = 1.
\label{elements}
\end{equation}
Equations~(\ref{elements}) and~(\ref{constraint}) reduce the number of generators of G($2$) to $14$. 
 
The dimensions of the fundamental and the adjoint representations of G($2$) are $7$ and $14$, respectively. Since the rank of the group is $2$, like the SU($3$) gauge group, only two of the generators can be diagonalized simultaneously.
It should be noted that all representations of G($2$) are real and therefore the seven-dimensional representation is equivalent to its complex conjugate. As a result, quarks and antiquarks in G($2$) gauge theory are indistinguishable.

In G($2$), the decomposition of the tensor product of three adjoint representations contains the fundamental representation, i.e.,
\begin{equation}
\label{product14}
\{14\} \otimes \{14\} \otimes \{14\}=\{7\} \oplus ... \ .
\end{equation}
As a consequence, three G($2$) gluons can screen a single G($2$) quark, i.e.,
\begin{equation}
\label{product7}
\{7\} \otimes \{7\}=\{1\} \oplus ...  \ .
\end{equation}
Therefore unlike SU($N$), even the seven-dimensional fundamental representation of G($2$) can be
screened by a bunch of gluons.

It is interesting to look at the homotopy groups because
they tell us what kind of topological excitations can arise.
Center vortices, monopoles, and instantons are classified
according to the $1$th, $2$th, and $3$th homotopy groups. The first homotopy group 
\begin{equation}
  \pi_1(G(2)/\mathbb I) = \mathbb I.
\end{equation}
It implies that  center vortices are absent in G($2$) theories, while for SU($2$) and SU($3$)
\begin{equation}
  \pi_1(SU(3)/\mathbb Z_3) = \mathbb Z_3,~~~~~~\pi_1(SU(2)/\mathbb Z_2) = \mathbb Z_2,
\end{equation}
which means that center vortices are present in these theories.
In the above homotopy groups, $\mathbb I$, $\mathbb Z_2$, and $\mathbb Z_3$ are center groups of G($2$), SU($2$), and SU($3$), respectively. It is clear that the center of G($2$) is trivial and that is why there are no vortices in G($2$) gauge theories.

\begin{figure}
\begin{center}
\resizebox{0.87\textwidth}{!}{
\includegraphics{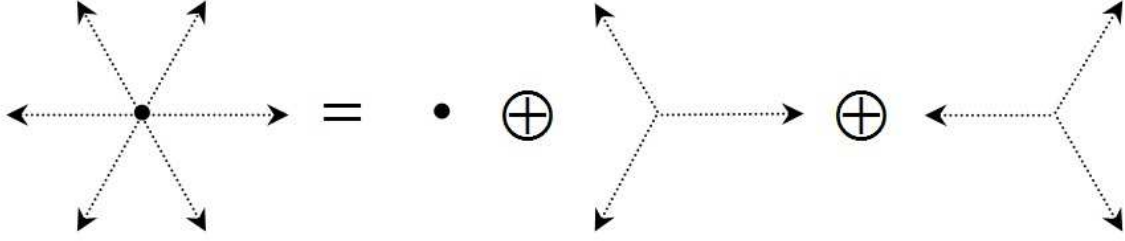}}
\caption{\label{pot}
The weight diagram for the seven-dimensional representation of the G($2$) group, decomposed to the
weight diagrams of the representations of the SU($3$) gauge group.}
\end{center}
\end{figure}

\begin{figure}
\begin{center}
\resizebox{0.87\textwidth}{!}{
\includegraphics{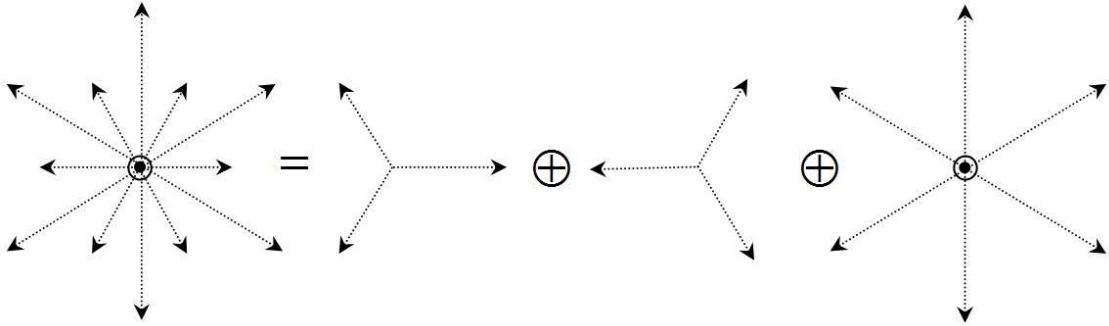}}
\caption{\label{po}
The weight diagram for the $14$-dimensional representation of the G($2$) group, decomposed to the
weight diagrams of the representations of SU($3$) gauge group.}
\end{center}
\end{figure}

The entire G($2$) group can be covered by six SU($2$) subgroups \cite{Cossum2007}:
\begin{eqnarray}
\begin{array}{llll}
1.~ H_1, & H_2, 
&  H_3
 \\
2.~H_4, & H_5, 
&  \frac{1}{2} \left( \sqrt{3} H_8 + H_3 \right)
 \\
3.~H_6, & H_7, 
&  \frac{1}{2} \left( - \sqrt{3} H_8 + H_3 \right) 
 \\
4.~\sqrt{3} H_8, & \sqrt{3} H_{9}, 
&  \sqrt{3} H_{10}
 \\
5.~\sqrt{3} H_{11}, & \sqrt{3} H_{12}, 
&  \frac{1}{2} \left( - \sqrt{3} H_8 + 3 H_3 \right)
\\
6.~\sqrt{3} H_{13}, & \sqrt{3} H_{14}, 
&  \frac{1}{2} \left( \sqrt{3} H_8 + 3 H_3 \right)
\label{gen}
\end{array} \, ,
\end{eqnarray}
where ${H}_{3}$ and ${H}_{8}$ are Cartan generators. The first three SU($2$) subgroups form four-dimensional real representations which are nonreducible. They generate an SU($3$) subgroup of G($2$) which is seven-dimensional and reducible. 
The representations of the remaining three SU($2$) subgroups are seven-dimensional, but they are reducible.
The decomposed weight diagrams of the fundamental and adjoint representations of the G($2$) gauge group into the weight diagrams of the SU($3$) representations are shown in Figs. 
\ref{pot} and \ref{po}, respectively. Since the weight diagram of the SU($3$) fundamental representation is two-dimensional, the weight diagrams of the SU($3$) subgroup of G($2$) is also two-dimensional and members of a multiplet correspond to  points in a plane.  
Therefore, under SU($3$) subgroup transformations, the seven- and $14$-dimensional representations decompose into
\begin{equation}
\label{7to3}
\{7\} = \{3\} \oplus \{\overline 3\} \oplus \{1\} ,
\end{equation}
\begin{equation}
\label{14to3}
\{14\} = \{8\} \oplus \{3\} \oplus \{\overline 3\}.
\end{equation}

It means that fourteen G($2$) gluons can be constructed from eight SU($3$) gluons plus six
additional gluons which are like the SU($3$) fundamental quark and antiquark. One of the differences between
the six gluons and the SU($3$) quarks is that the former ones are bosons while the latter ones are fermions. 
\underline{}
The Cartan generators of the G($2$) gauge group can be constructed by the SU($3$) Cartan generators,
\begin{equation}
\label{Cartan 1}
H_a^{7} = \frac{1}{\sqrt{2}} \left( \begin{array}{ccc} \lambda_a^3 &0 & 0 \\
0 & \; 0 & 0\\ 0 & 0 & -(\lambda_a^3)^* \ \end{array}\right),~~~~~~H_a^{14} = \frac{1}{\sqrt{8}} \left( \begin{array}{ccc} \lambda_a^3 &0 & 0 \\
0 & \; -({\lambda_a^3})^* & 0\\ 0 & 0 & \lambda_a^8\ 
\end{array} \right),
\end{equation}
where $\lambda_a^3$ and $\lambda_a^8$ ($a=3,8$) are the SU($3$) Cartan generators in the fundamental and adjoint representations, respectively. 
For all representations of G($2$), we use the following normalization condition for the generators:
\begin{equation}
\label{normalize}
\mathrm{Tr}[T_{a}T_{b}]=\frac{1}{2}\delta_{ab}.
\end{equation}

In the SU($3$) subgroup of G($2$), the center elements of G($2$) can be constructed from the group $\mathbb Z_3$, the center of SU($3$): 
\begin{equation}
\label{Z}
Z^{7}= \left(\begin{array}{ccc} zI_{3\times3} & 0 & 0 \\0 &1 &0 \\ 0 & 0 & z^*I_{3\times3}
\end{array} \right),~~~~~~Z^{14}= \left(\begin{array}{ccc} zI_{3\times3} & 0 & 0 \\0 &z^*I_{3\times3} &0 \\ 0 & 0 & I_{8\times8}
\end{array} \right),
\end{equation}
where $I$ is the unit matrix and $z\in\{1,e^{\pm\frac{2\pi i}{3}}\}$ is an element of $\mathbb Z_3$, the center group of SU($3$).

The center element in various representations of SU($N$) is equal to $z^{k_r}$, where $k_r$ is the $N$-ality of $r$-dimensional representation.
 The $N$-ality classifies representations of the SU($N$) group with respect to the center group, $\mathbb Z_N$. $3$-ality of the fundamental and adjoint representations of the SU($3$) gauge group are equal to $1$ and $0$, respectively. Therefore, in Eq.~(\ref{Z}), $zI_{3\times3}$, $z^*I_{3\times3}$ and $I_{8\times8}$ are center elements of fundamental, its complex conjugate and adjoint representations of SU($3$) group. The number $1$ corresponds to the one-dimensional representation in Eq.~(\ref{7to3}).

\begin{figure}
\begin{center}
\resizebox{0.87\textwidth}{!}{
\includegraphics{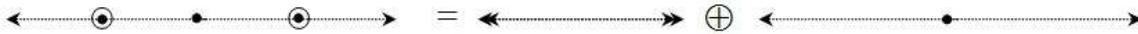}}
\caption{\label{po1}
The weight diagram for the seven-dimensional representation of the SU($2$) subgroup of G($2$), decomposed to the
weight diagrams of the representations of the SU($2$) gauge group.} 
\end{center}
\end{figure}

\begin{figure}
\begin{center}
\resizebox{0.87\textwidth}{!}{
\includegraphics{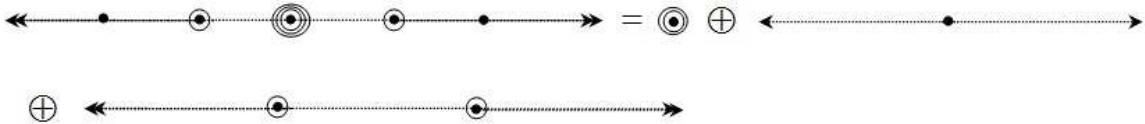}}
\caption{\label{po2}
The weight diagram for the $14$-dimensional representation of the SU($2$) subgroup of G($2$), decomposed to the
weight diagrams of the representations of the SU($2$) gauge group.} 
\end{center}
\end{figure}

In addition to the SU($3$) subgroup, the second three SU($2$) subgroups in Eq.~(\ref{gen}) are seven-dimensional but they are reducible.
The weight diagrams of the fundamental and adjoint representations of SU($2$) subgroup of G($2$) gauge group decomposed into the weight diagrams of the SU($2$) gauge group 
are shown in Figs. \ref{po1} and \ref{po2}, respectively. 
Since the weight diagram of SU($2$) fundamental representation is one-dimensional,
 the weight diagrams of the SU($2$) subgroup are also one-dimensional and the members of a multiplet correspond to points along a line. Therefore, under SU($2$) subgroup transformations, the seven- and $14$-dimensional representations decompose into (see the Appendix for details)
\begin{equation}
\label{7t0223}
\{7\} = 2\{2\} \oplus \{3\} ,
\end{equation}
\begin{equation}
\label{14to31324}
\{14\} = 3\{1\} \oplus \{3\} \oplus 2\{4\}
\end{equation}
where the two- and three-dimensional representations are the fundamental and adjoint representations of SU($2$), respectively.
The Cartan generator ${H}_{8}$ in the SU($2$) subgroup with generators ${H}_{8}$, ${H}_{9}$, and ${H}_{10}$ are given by
\begin{equation}
\label{Cartan 2}
H_8^{7} = \frac{1}{\sqrt{6}} \left( \begin{array}{ccc} \sigma_3^2 &0 & 0 \\
0 & \sigma_3^2 & 0 \\
0 & 0 & \sigma_3^3 \ \end{array}\right),~~~~~~H_8^{14} = \frac{1}{\sqrt{24}} \left( \begin{array}{cccccc} \sigma_3^3 &0 &0 &0 &0 & 0 \\
 0 &\sigma_3^4 &0 &0 &0 & 0\\0& 0& \sigma_3^4\ &0 &0 &0\\ 0 & 0 & 0 & 0 & 0 & 0\\ 0 & 0 & 0 & 0 &0 & 0\\ 0 &0 & 0 & 0 & 0&0\
\end{array} \right)
\end{equation}
where $\sigma_3^2$, $\sigma_3^3$, and $\sigma_3^4$ are the Cartans of SU($2$) in the fundamental, adjoint and, four-dimensional representations, respectively.
The center elements of the SU($2$) subgroup are given by 
\begin{equation}
\label{ZZ}
Z^{7}= \left(\begin{array}{ccc} zI_{2\times2} & 0 & 0 \\0 &zI_{2\times2} &0 \\ 0 & 0 & I_{3\times3}
\end{array} \right),~~~~~~Z^{14}= \left(\begin{array}{cccccc} I_{3\times3} & 0 & 0&0&0&0 \\0 &zI_{4\times4} &0 &0&0&0\\ 0 & 0 & zI_{4\times4}&0&0&0\\0&0&0&1&0&0\\0&0&0&0&1&0\\0&0&0&0&0&1
\end{array} \right)
\end{equation}
where $I$ is the unit matrix and $z\in\{1,e^{\pi i}\}$ is an element of $\mathbb Z_2$, center group of SU($2$).
The SU($2$) representations are labeled by a spin index $j$ with either half-integer or integer values. $2$-ality is zero for all integer $j$ representations of the SU($2$) gauge group and $1$ for half-integer ones i.e. the $2$-ality of two-dimensional ($j=1/2$), three-dimensional ($j=1$) and four-dimensional ($j=3/2$) representations are $1$, $0$, and $1$, respectively. Therefore, $zI_{2\times2}$, $I_{3\times3}$ and $zI_{4\times4}$ are center elements in two-, three-, and four-dimensional representations of the SU($2$) gauge group, respectively, and $1$ corresponds to a representation with  dimension $1$.

\section{CONFINEMENT IN THE G($2$) GAUGE GROUP}

\begin{figure}
\begin{center}
\vspace{90pt}
\resizebox{0.6\textwidth}{!}{
\includegraphics{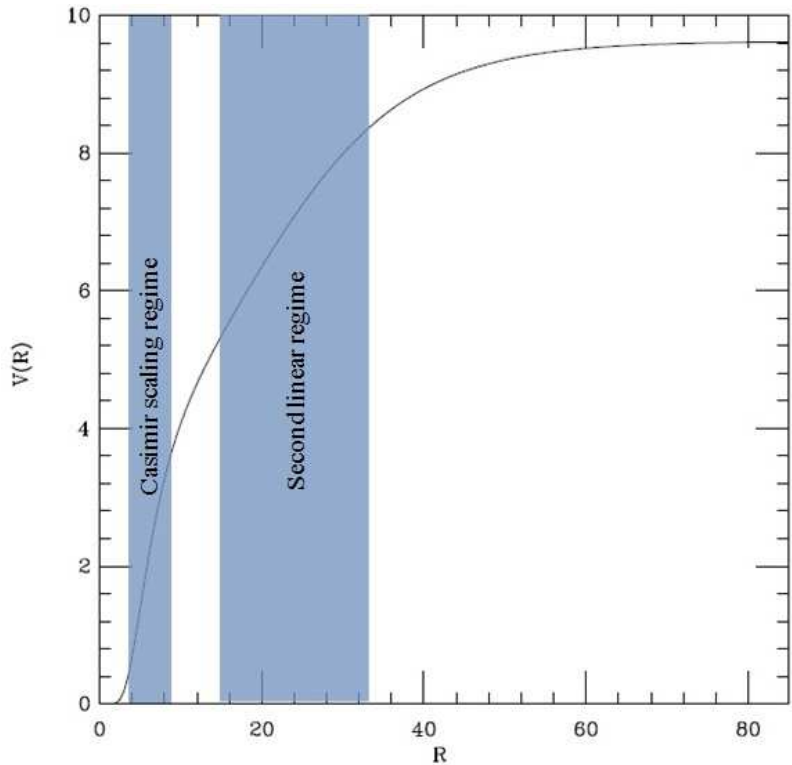}}
\caption{\label{MM}
It seems that there are two linear regimes for G($2$) gauge group. The first one agrees qualitatively with the Casimir scaling \cite{Deldar2012}.  
}
\end{center}
\end{figure}

The center vortex model is able to describe confinement for a gauge group with nontrivial center elements. Confinement is obtained from random fluctuations in the number of center vortices which link to the Wilson loops. Therefore, no center vortices means no confinement and the linear regime should not be observed in gauge theories, such as G($2$) without nontrivial center elements. However, numerical lattice calculations show confinement for the G($2$) gauge theory. 
In our previous calculations \cite{Deldar2012}, we applied the domain model to the G($2$) gauge group which has only a trivial center element $I$. The potential energy, $V(R)$, between two heavy sources in seven-dimensional representation is plotted in Fig. \ref{MM}. When the domain is completely contained within the Wilson loop area, Eq. (\ref {alfa1}) for G($2$) is  
\begin{equation}
\label{alfa2}
\exp(i\vec{\alpha}^{(0)}\cdot\vec{H})= I.
\end{equation}
At large distances where the vacuum domain is located completely inside the Wilson loop, the string tension is zero. This is because the 
total magnetic flux which is carried by the domain is zero and it has no effect on the loop. At intermediate distances, a linear regime is observed. For this regime, the vacuum domain is partially located inside the Wilson loop. Therefore a nonzero magnetic flux of the vacuum domain is located inside the 
Wilson loop which leads to a nonzero string tension.
It seems that the linear regime of G($2$), from the onset of confinement
to the onset of color screening, has two different slopes. String tension of the first one (in lower energy compared with the second one) is proportional to the eigenvalue of the
 quadratic Casimir operator of the corresponding representation. In other words, the first linear regime is qualitatively in agreement with Casimir scaling.
The result from our previous paper is 
 \begin{equation}
\frac{K_{14}}{K_{f}}=1.48 ~~~~~~~~~~~~~~~~~~~~~~~\frac{K_{27}}{K_{f}}=1.65
\label{string ratio}
\end{equation}
while the Casimir ratios are
\begin{equation}
\frac{C_{14}}{C_{f}}=2 ~~~~~~~~~~~~~~~~~~~~~~~\frac{C_{27}}{C_{f}}=\frac{7}{3}.
\end{equation}
The string ratios are qualitatively in rough agreement with Casimir ratios.  

To study the reasons of confinement in G($2$) gauge group, we apply the thick center vortex model to the subgroups of G($2$).   

\subsection{SU($3$) subgroup of G($2$)}
First, we study the SU($3$) subgroup of G($2$). The potential between two heavy sources is given by
\begin{equation}
\label{SU(3)potential}
V(R) = \sum_{x}\ln\left\{ 1 - f_{1}(1 - {\mathrm {Re}} g_{r}
[\vec{\alpha}^1_{C}(x)])\right\}
\end{equation}
where $f_{1}$ is the probability that any given unit is pierced by a center vortex, $g_{r}$ has the same form as in Eq. (\ref{gr1}) and we use the flux profile:
\begin{equation}
\alpha_i^n(x)=\frac{\alpha_i^{n(max)}}{2}[1-\tanh(ay(x)+\frac{b}{R})],
\label{alpha}
\end{equation}
where $n$ indicates the domain type and $a , b$ are free parameters of the model, and $y(x)$ is
\begin{equation}
y(x)=\begin{cases} -x & \lvert R-x\lvert>x \\x-R & \lvert R-x\lvert\leqslant x
\end{cases}
\end{equation}
$y(x)$ is the nearest distance of $x$ from the timelike side of the loop and $\alpha_i^{n(max)}$ is the maximum value of the flux profile.
At large distances, where the vortex is completely inside the Wilson loop, 
we normalize $\exp(i\vec{\alpha}^n \cdot \vec{H})$ to the center element of the SU($3$) subgroup to study the role of the SU($3$) gauge group in 
observing G($2$) confinement, 
\begin{equation}
 \exp(i\vec{\alpha}^{max} \cdot \vec{H}^{f})=\exp(\alpha_1^{max}H_3^{f}+\alpha_2^{max}H_8^{f})=
  \left(\begin{array}{ccc} zI_{3\times3} & 0 & 0 \\0 &1 &0 \\ 0 & 0 & z^*I_{3\times3}
\end{array} \right), 
\label{expalpha3}
\end{equation}
where ${H}_a^{f}$ ($a=3,8$) are the Cartan generators in the seven-dimensional representation.  
We use the Cartan generators of the SU($3$) subgroup in the fundamental representation from Eq.~(\ref{Cartan 1}). The maximum values of flux profiles $\alpha_1^{max}$ and $\alpha_2^{max}$ for the fundamental representation are obtained:
\begin{equation}
\begin{array}{c}
\frac{2\pi}{3}=\frac{\alpha_1^{max}}{\sqrt{8}}+\frac{\alpha_2^{max}}{\sqrt{24}}
\\\frac{2\pi}{3}=\frac{-\alpha_1^{max}}{\sqrt{8}}+\frac{\alpha_2^{max}}{\sqrt{24}}
\end{array}\}
\Longrightarrow ~~\alpha_1^{max}=0, ~~\alpha_2^{max}=\frac{2\pi \sqrt{24}}{3}.
\label{exalpha}
\end{equation}
\begin{figure}
\begin{center}
\vspace{90pt}
\resizebox{.75\textwidth}{!}{
\includegraphics{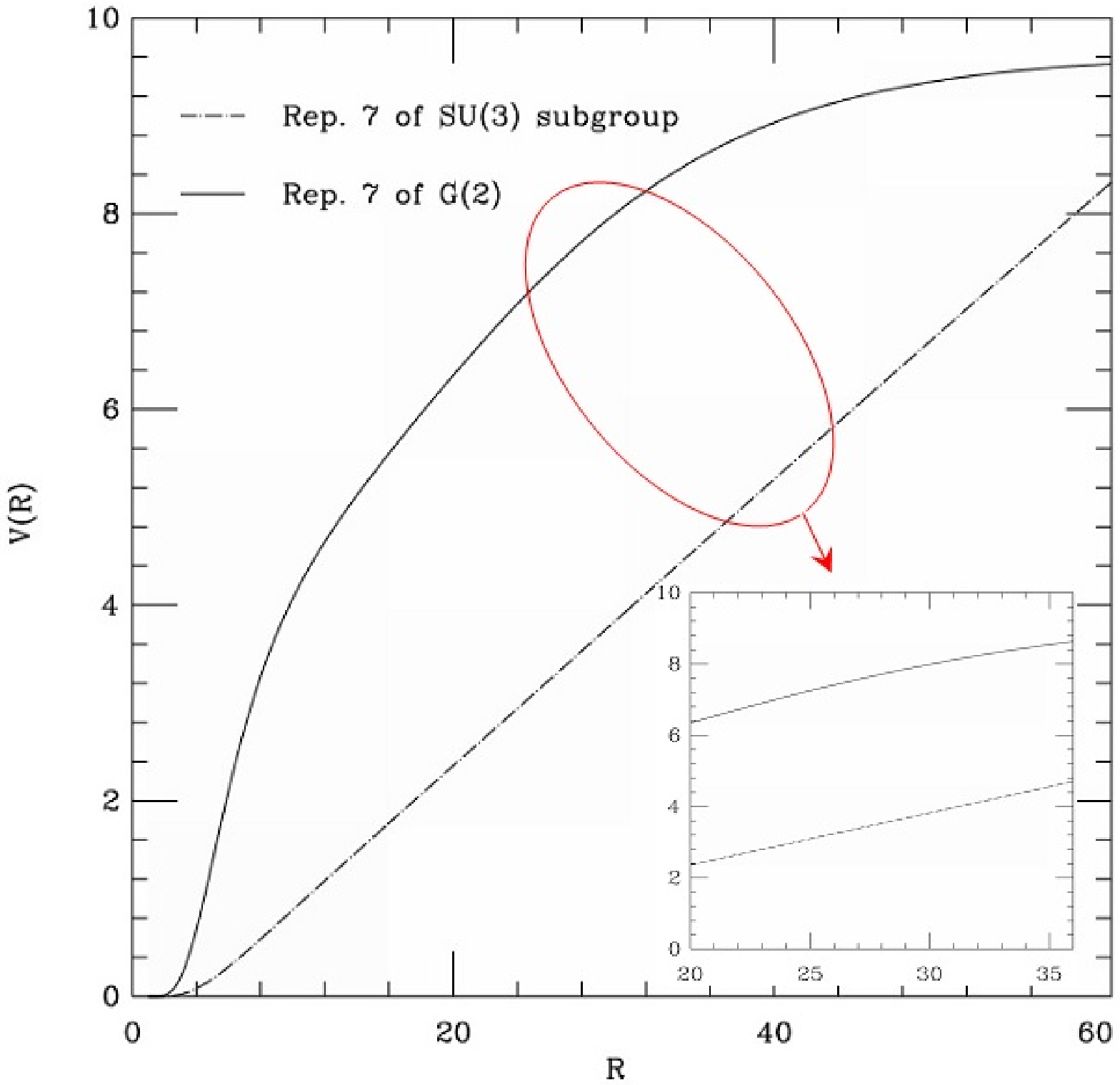}}
\caption{\label{S}
For $25 <R <33$, the slopes of the seven-dimensional representation potentials in G($2$) and its SU($3$) subgroups are roughly equal, 
in other words, the linear parts of the potentials are parallel in this regime.
}
\end{center}
\end{figure}
\begin{figure}
\begin{center}
\vspace{90pt}
\resizebox{0.8\textwidth}{!}{
\includegraphics{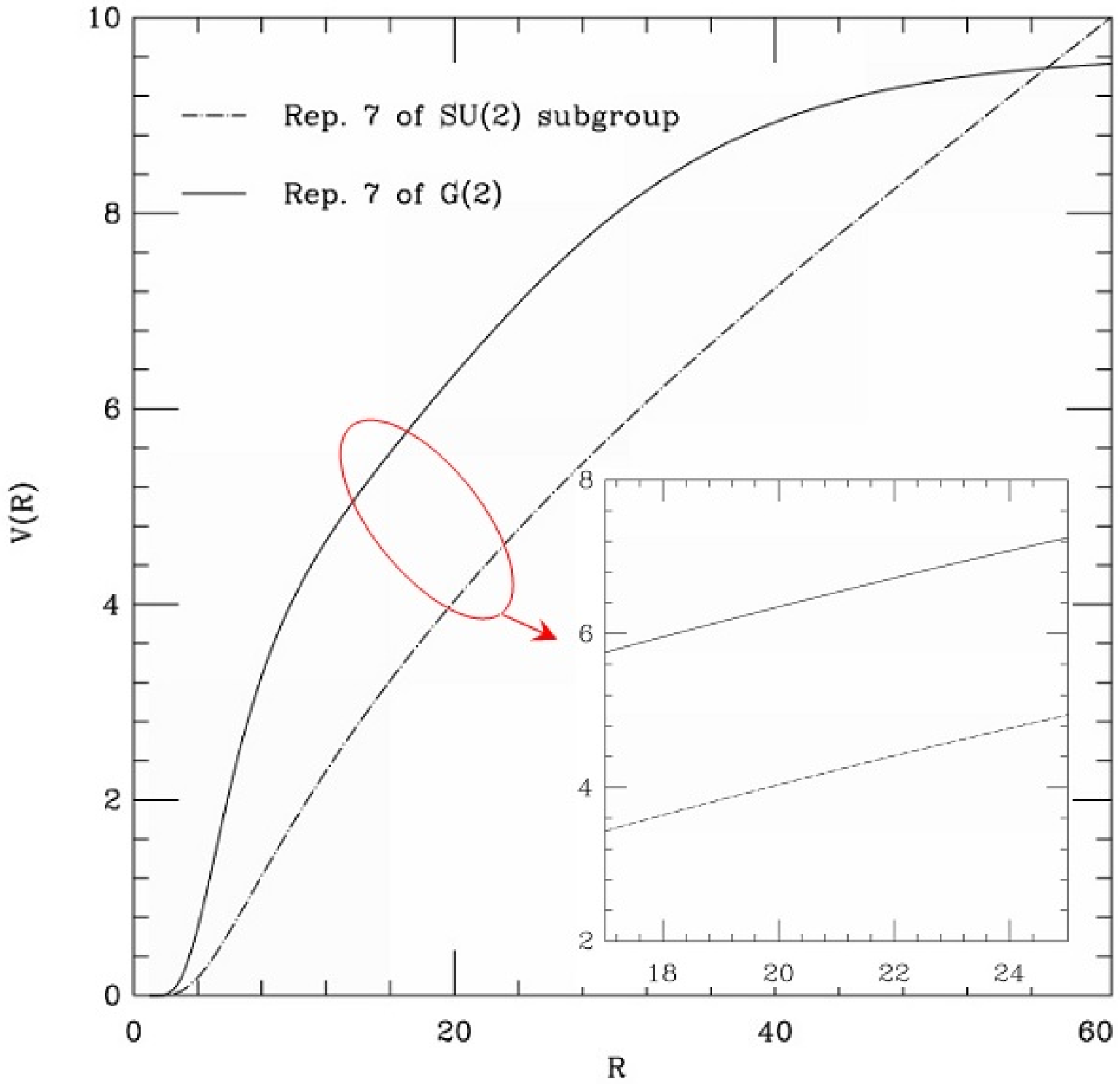}}
\caption{\label{S1}
For $17 <R <23$,  the slopes of the seven-dimensional representation potentials in G($2$) and SU($2$) subgroup are roughly equal. 
}
\end{center}
\end{figure}

Now, we are ready to calculate the potential from Eq.~(\ref{SU(3)potential}) by using the flux profile of Eq.~(\ref{alpha}). 
Figure \ref{S} plots the potentials for the fundamental representation in the SU($3$) subgroup and G($2$). The free parameters $a$, $b$, and $f_{1}$ are chosen to be $0.05$, $4$, and $0.1$,
 respectively.  $f_{1}$ is chosen to be  equal to $f_{0}$, the probability  that any given unit area is pierced by a vacuum domain in G($2$) gauge group.   As Fig. \ref{S} shows, 
the slope of the second linear regime  
of the fundamental representation in the G($2$) gauge group is roughly equal to the slope of the potential between 
static sources in its SU($3$) subgroup in $25 <R <33$. In percent for this interval, the difference between the slopes is not more than 7\%.
 
\subsection{SU($2$) subgroup of G($2$)}

Next, we apply the model to the SU($2$) subgroup of G($2$). 
We use the SU($2$) subgroup with the generators ${H}_{8}$, ${H}_{9}$, and ${H}_{10}$ which are among 
the second three SU($2$) subgroups in the list of the six SU($2$) subgroups in Eq.~(\ref{gen}) that cover the entire G($2$) group. The Cartan generator ${H}_{8}$ is given in
 Eq.~(\ref{Cartan 2}). 
 The potential between two static sources is obtained from Eq.~(\ref{SU(3)potential}).
 This time, to investigate the role of SU($2$) in the confinement of G($2$) quarks, we normalize $\exp(i\vec{\alpha}^n \cdot \vec{H})$ to the center elements of the SU($2$) subgroup, 
\begin{equation}
 \exp(i\vec{\alpha}^{max} \cdot \vec{H}^{f})=\exp(\alpha^{max}H_8^{f})=
  \left(\begin{array}{ccc} zI_{2\times2} & 0 & 0 \\0 &zI_{2\times2} &0 \\ 0 & 0 & I_{3\times3}
\end{array} \right). 
\label{expalpha2}
\end{equation}
Thus, the maximum value of the flux profile $\alpha^{max}$ for the fundamental representation is obtained:
\begin{equation}
 \alpha^{max}=2\pi{\sqrt{6}}.
\label{expalpha}
\end{equation}
To compare the role of the SU($2$) gauge group in confining quarks in the G($2$) gauge group, we plot the
 potential between two G($2$) fundamental heavy quarks and two heavy quarks in its SU($2$) subgroup in Fig. \ref{S1}.  
As Fig. \ref{S1} shows, the slope of the second linear regime  
of the fundamental representation in the G($2$) gauge group is roughly equal to the slope of the potential between static sources in its SU($2$) subgroup in $17 <R <23$.
 In this interval, the difference between the slopes is not more than 3\%.

\subsection{Adjoint representation}

\begin{figure}
\begin{center}
\vspace{90pt}
\resizebox{0.78\textwidth}{!}{
\includegraphics{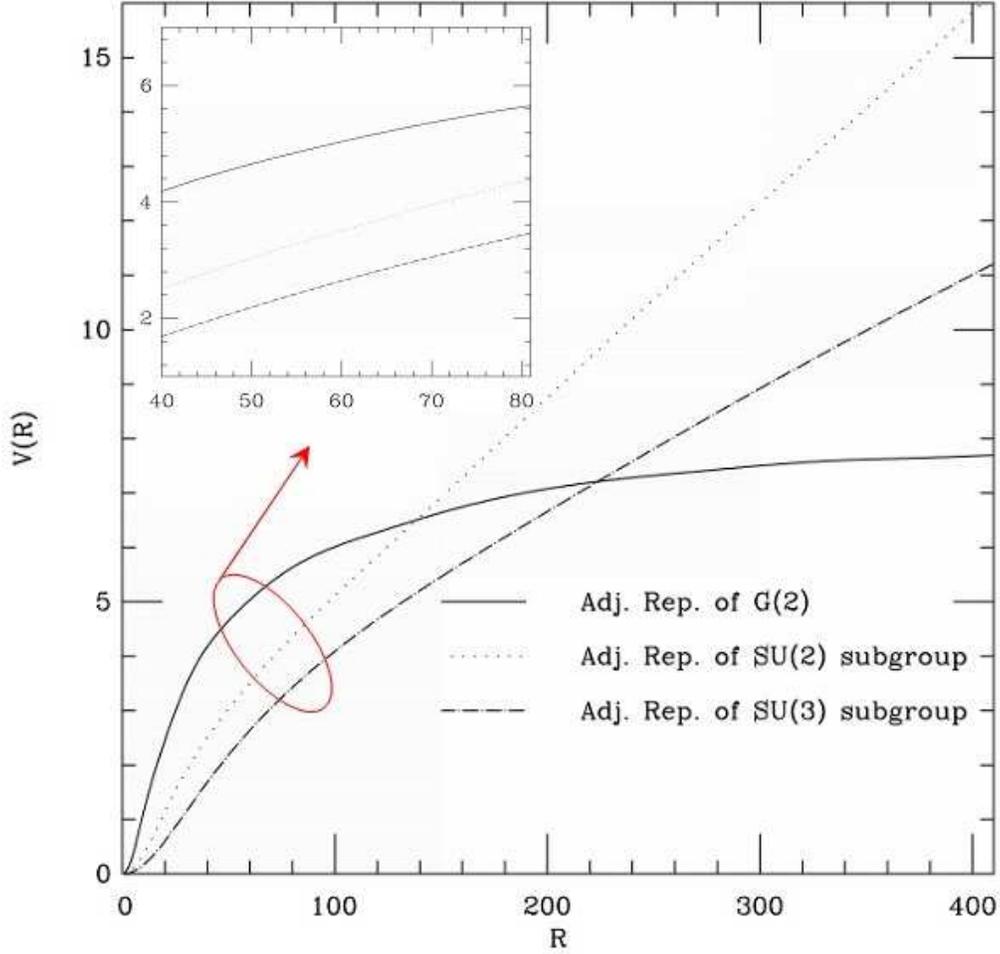}}
\caption{\label{adjoint}
For $40 <R <80$,  the slopes of the $14$-dimensional representation potentials for G($2$), SU($2$) and  SU($3$) subgroups are roughly equal. 
}
\end{center}
\end{figure}

To study the role of G($2$) subgroups in observing the linear potential, we obtain the potentials for the static sources in the adjoint 
representation for G($2$) and its subgroups. From Eqs. 
(\ref{Z}) and (\ref{ZZ}), one sees that the adjoint G($2$) contains SU($2$) and SU($3$) nontrivial $N$-alities, hence their potentials must be nontrivial. 
We use Eq. (\ref{SU(3)potential}) and the flux profile of Ref. \cite{Deldar2010}. 
For the adjoint representation of the G($2$) gauge group, when the domain is completely contained within the Wilson loop,
\begin{equation}
\label{alfa220}
\exp(i\vec{\alpha}^{(0)}\cdot\vec{H_{adj}})= I.
\end{equation}  
Therefore, the maximum values of the flux profiles $\alpha_1^{max}$ 
and $\alpha_2^{max}$ for the adjoint representation are zero and $4\pi \sqrt{24}$.  Using Eqs. (\ref{Z}) and (\ref{ZZ}) in Eq. (\ref {alfa1}),
the maximum values of the flux profiles for the SU($2$) and SU($3$) subgroups are $2\pi{\sqrt{24}}$ and $\frac{4\pi \sqrt{24}}{3}$,  respectively.
Figure \ref{adjoint} plots the potentials between two color sources in the adjoint representation of the G($2$) gauge group and its subgroups.
Like the fundamental representation, there is a regime where the potentials have roughly the same slope. 

Considering the results of subsections A, B, and C, one can argue that the SU($2$) and SU($3$) subgroups have dominant roles in the second linear 
regime. The nonzero 
flux profile of $\alpha(x)$ in the G($2$), SU($3$) and SU($2$) corresponds  to ${H}_{8}$. 

 The results are in agreement with lattice results. Greensite $\it{et~ al.}$ \cite{Greensite2007} have studied different projected loops in lattice gauge theory. The slopes calculated from
 the Wilson loops of the SU($2$)-only and SU($3$)-only link
variables are close to the G($2$) full Wilson loops. SU($3$) and SU($2$) removed loops give slopes different from the full G($2$) Wilson loops.

On the other hand, Pepe $\it{et~ al.}$ \cite{Holland2003} have studied the G($2$) gauge-Higgs theory where G($2$) has been
spontaneously broken to the SU($3$) gauge theory by adding the Higgs field in the seven-dimensional representation to the Lagrangian density of G($2$). 
As a result, the Higgs field gives a mass to $6$ of the $14$ gluons (the mass of six gluons is proportional to the expectation value of the Higgs field)
 while remaining gluons, associated with the SU($3$) subgroup, remain massless. Therefore gauge-Higgs theory can interpolate between pure the SU($3$) 
subgroup of G($2$) and pure G($2$).

\section{More on SU($2$) and SU($3$) subgroups}
 
\begin{figure}
\begin{center}
\vspace{90pt}
\resizebox{0.6\textwidth}{!}{
\includegraphics{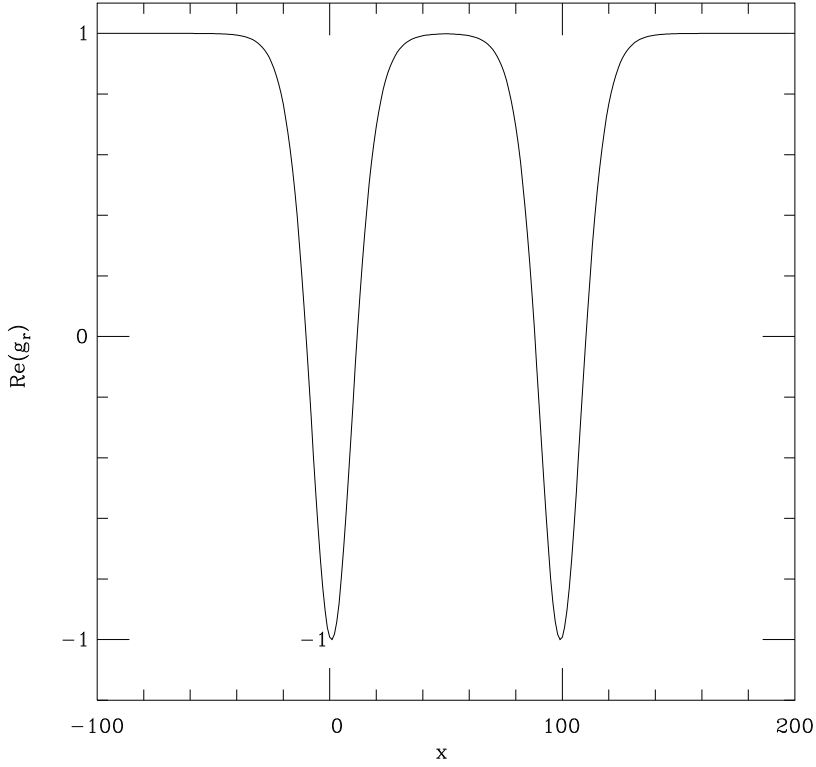}}
\vspace{-40pt}
\caption{\label{SSS}
 ${\mathrm {Re}}(g_{r})$ versus $x$ is plotted for $R = 100$ in the two-dimensional representation ($j=1/2$) of the SU($2$) gauge group. ${min}[{\mathrm {Re}g_{r}}]=-1$ corresponds to the
nontrivial center element of  SU($2$).}
\end{center}
\end{figure}

In Sec. IV we have shown that the slope of the potentials for the G($2$) and its SU($2$) and SU($3$) subgroups are equal for an interval at intermediate distances.
In this section, we study the vortex profile to confirm that the subgroups of G($2$)  have some important roles in confining G($2$) heavy sources at intermediate distances.
In general,  for SU($N$) gauge theory, ${\mathrm {Re}}(g_{r})$ changes from $1$ to the values which correspond
to the nontrivial center elements. This is true even if we use only the vacuum domains.
Using the domain model, ${\mathrm {Re}}(g_{r})=1$ when either the trivial center domain is entirely contained within the Wilson loop
or the domains are far outside the Wilson loop. For the G($2$) gauge group, ${\mathrm {Re}}(g_{r})$ changes from $1$ to some values which we are going to explain in 
terms of the G($2$) subgroups.
First, we calculate ${\mathrm {Re}}(g_{r})$  in SU($2$) gauge theory, using only the trivial center element. From Eq. (\ref{gr1}) and the Cartan of the SU($2$) gauge group, 
${\mathrm {Re}}(g_{r})$ in the fundamental representation of SU($2$) is obtained:
 \begin{equation}
 {\mathrm {Re}}(g_{j=1/2})=cos(\frac{\alpha}{2})=\frac {sin(\alpha)}{2sin(\frac{\alpha}{2})}. 
 \end{equation}
The flux profile is the same as Eq. (\ref{alpha}) where $\alpha^{max}=4\pi$. The free parameters $a$ and $b$ are chosen to be $0.05$ and $4$, 
respectively. 
Figure \ref{SSS} plots ${\mathrm {Re}}(g_{r})$ versus $x$ for $R = 100$ in the fundamental representation of SU($2$). $x$ shows the location of 
the domain. The left and right legs of the Wilson loop are located at zero and $x = R$, respectively. The size of each domain is proportional 
to the inverse of $a$. With our chosen parameters, the size of the vacuum domain is about $20$. Since the  domain  locates completely
 inside the Wilson loop, when the spatial length is equal to $100$, then  ${\mathrm {Re}}(g_{r})$ is equal to $1$ for the interval $[20,80]$.
${\mathrm {Re}}(g_{r})$ changes between the two values: $1$, when the vacuum domain is located completely inside the loop, and $-1$. The value 
of  $-1$ corresponds to the value of the 
nontrivial center element of SU($2$) gauge group:
\begin{equation}
 { {min}[\mathrm{Re}}(g_{r})]={\mathrm {Re}}(e^{i\pi})=-1.
\end{equation}
$e^{i\pi}$ comes from the nontrivial center element of the SU($2$) group. 
 
\begin{figure}
\begin{center}
\vspace{90pt}
\resizebox{0.6\textwidth}{!}{
\includegraphics{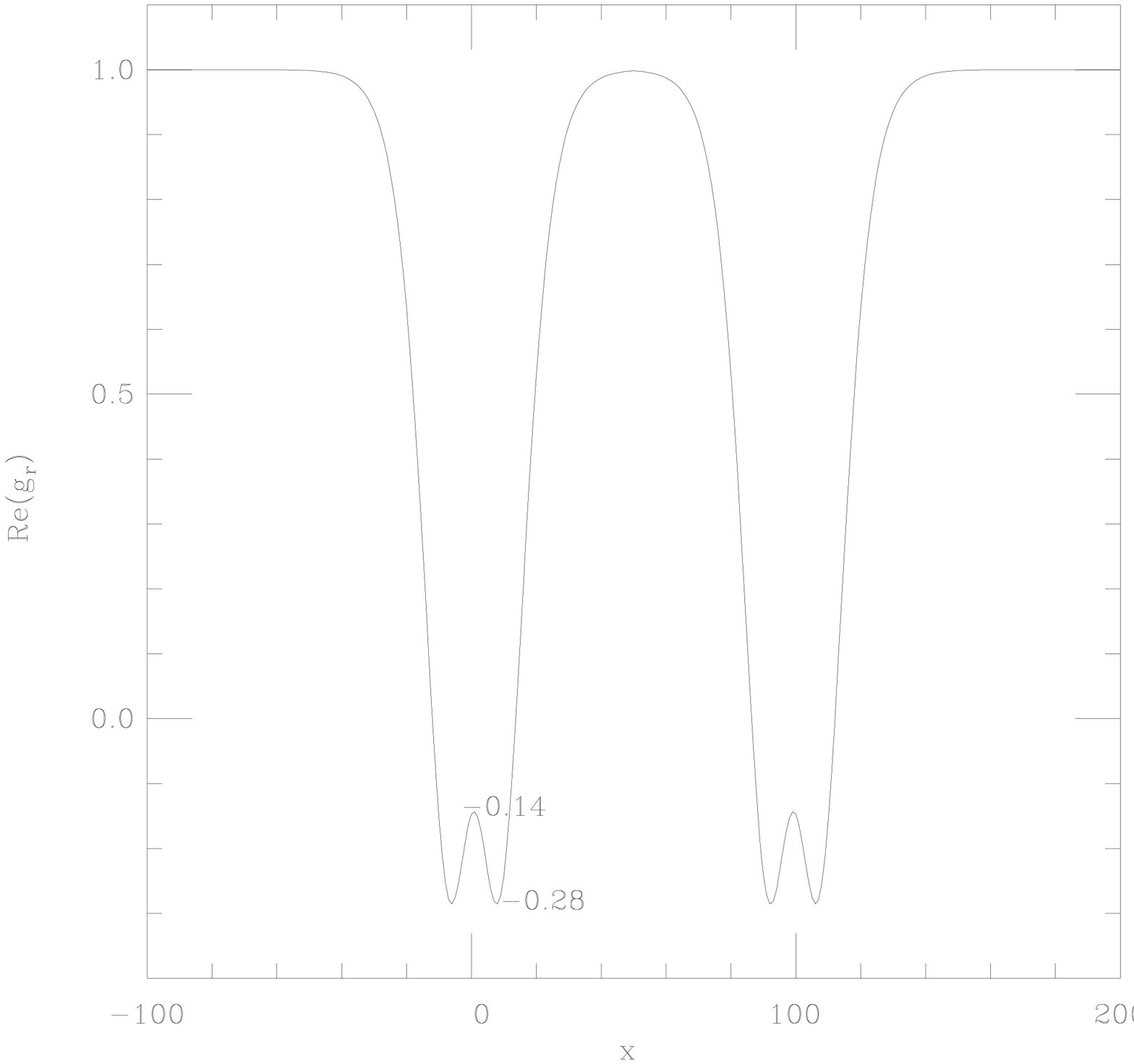}}
\vspace{-40pt}
\caption{\label{ggg}
  ${\mathrm {Re}}(g_{r})$ versus $x$ is plotted for $R = 100$ in the seven-dimensional representation of the G($2$) gauge group. 
${\mathrm {Re}}(g_{r})$ has a maximum value of $1$ and two extremums -$0.14$ and -$0.28$. When the  vacuum domain locates completely inside the 
Wilson loop, ${\mathrm {Re}}(g_{r})$ reaches to  $1$. The extremums are explained by the SU($2$) and the SU($3$) subgroups of G($2$). }
\end{center}
\end{figure}

G($2$) does not have any nontrivial center element. The upper limit of ${\mathrm {Re}}(g_{r})$ is $1$ but  
 the lower limits  are not known. We interpret the values of  the lower limits in terms of the SU($2$) and SU($3$) subgroups of the G($2$) gauge group. 
${\mathrm {Re}}(g_{r})$ in the fundamental representation of G($2$) can be obtained from Eqs. (\ref{gr1}) and (\ref{Cartan 1}) (left):
  \begin{equation}
{\mathrm {Re}}(g_{7})=\frac{1}{7}(4 cos(\frac{\alpha_2}{2\sqrt{6}})+2 cos(\frac{\alpha_2}{\sqrt{6}})+1)
  \end{equation}
 where the maximum flux profiles in the fundamental representation,  $\alpha_1^{max}=0$ and $\alpha_2^{max}={2\pi \sqrt{24}}$, are
obtained by using the Cartan generators of Eq. (\ref{Cartan 1}). 
Figure \ref{ggg} plots ${\mathrm {Re}}(g_{r})$ versus $x$ for $R = 100$ for the fundamental representation of G($2$). The flux profile and its free parameters are chosen as the same 
as the SU($2$) gauge group. As Fig. \ref{ggg} shows, the maximum of ${\mathrm {Re}}(g_{r})$ is equal to $1$ which is expected when the vacuum domain is located completely inside 
the loop. The interesting points are the extremums at -$0.28$ and -$0.14$.
 We try to explain these  extremums by the SU($2$) and SU($3$) subgroups of G($2$). If  the center vortices of the SU($2$) or the SU($3$) subgroup are 
located completely inside the  loop, then
\begin{equation}
\label{alfa11}
\exp(i\vec{\alpha}\cdot\vec{H})=Z_s
\end{equation} 
 where $Z_s$ represents the center elements of the subgroups of G($2$).

The minimum of the ${\mathrm {Re}}(g_{r})$ for the SU($2$) subgroup of the G($2$) gauge group for the fundamental representation  when
 the center elements are located completely inside the Wilson loop can be obtained from Eq.~(\ref{ZZ}) (left) with $z=e^{i\pi}$:
\begin{equation}
\begin{aligned}
 {min}[{\mathrm {Re}}g_{r}
(\alpha(x) )]_ {SU(2)}&=
\frac{1}{7}{\mathrm {Re}}(\mathrm{Tr}(e^{i\alpha \cdot
H})_\mathrm{min})=\frac{1}{7}{\mathrm {Re}}\mathrm{Tr}\left(\begin{array}{ccc} e^{i\pi}I_{2\times2} & 0 & 0 \\0 &e^{i\pi}I_{2\times2} &0 \\ 0 & 0 & I_{3\times3}
\end{array} \right)\\ 
&=\frac {1}{7}(-2-2+3)=-0.14.
\end{aligned}
\end{equation}
This value is equal with one of the  extremums of ${\mathrm {Re}}(g_{r})$ of the G($2$) gauge group in Fig. \ref{ggg}. The second extremum, -$0.28$, 
happens because of the SU($3$) subgroup which has been explained in our previous paper \cite{Deldar2012}. 
Using Eq. (\ref{Z}) (left) with $z=e^{\frac {i2\pi}{3}}$, the minimum of ${\mathrm {Re}}(g_{r})$ of the SU($3$) subgroup of G($2$) in the 
fundamental representation is obtained 
\begin{equation}
\begin{aligned}
 {min}[{\mathrm {Re}}g_{r}
(\alpha(x))]_ {SU(3)}&=\frac{1}{7}{\mathrm {Re}}(\mathrm{Tr}(e^{i\alpha \cdot
H})_\mathrm{min})=\frac{1}{7}{\mathrm {Re}}\mathrm{Tr}\left(\begin{array}{ccc} e^\frac{i2\pi}{3}I_{3\times3} & 0 & 0 \\0 &1 &0 \\ 0 & 0 & e^{-\frac {i2\pi}{3}}I_{3\times3}
\end{array} \right)\\
&=\frac {1}{7}(-1.5+1-1.5)=-0.28.
 \end{aligned}
\end{equation}   
 This value is equal to another extremum of ${\mathrm {Re}}(g_{r})$ of the G($2$) gauge group in Fig. \ref{ggg}.
Therefore, the extremums -$0.14$ and -$0.28$ in ${\mathrm {Re}}(g_{r})$ of the G($2$) in the fundamental representation correspond to the SU($2$) and SU($3$) subgroups.
However, the absolute minimum of ${\mathrm {Re}}(g_{r})$ of G($2$) corresponds to the SU($3$) subgroup. 
 
\begin{figure}
\begin{center}
\vspace{90pt}
\resizebox{0.6\textwidth}{!}{
\includegraphics{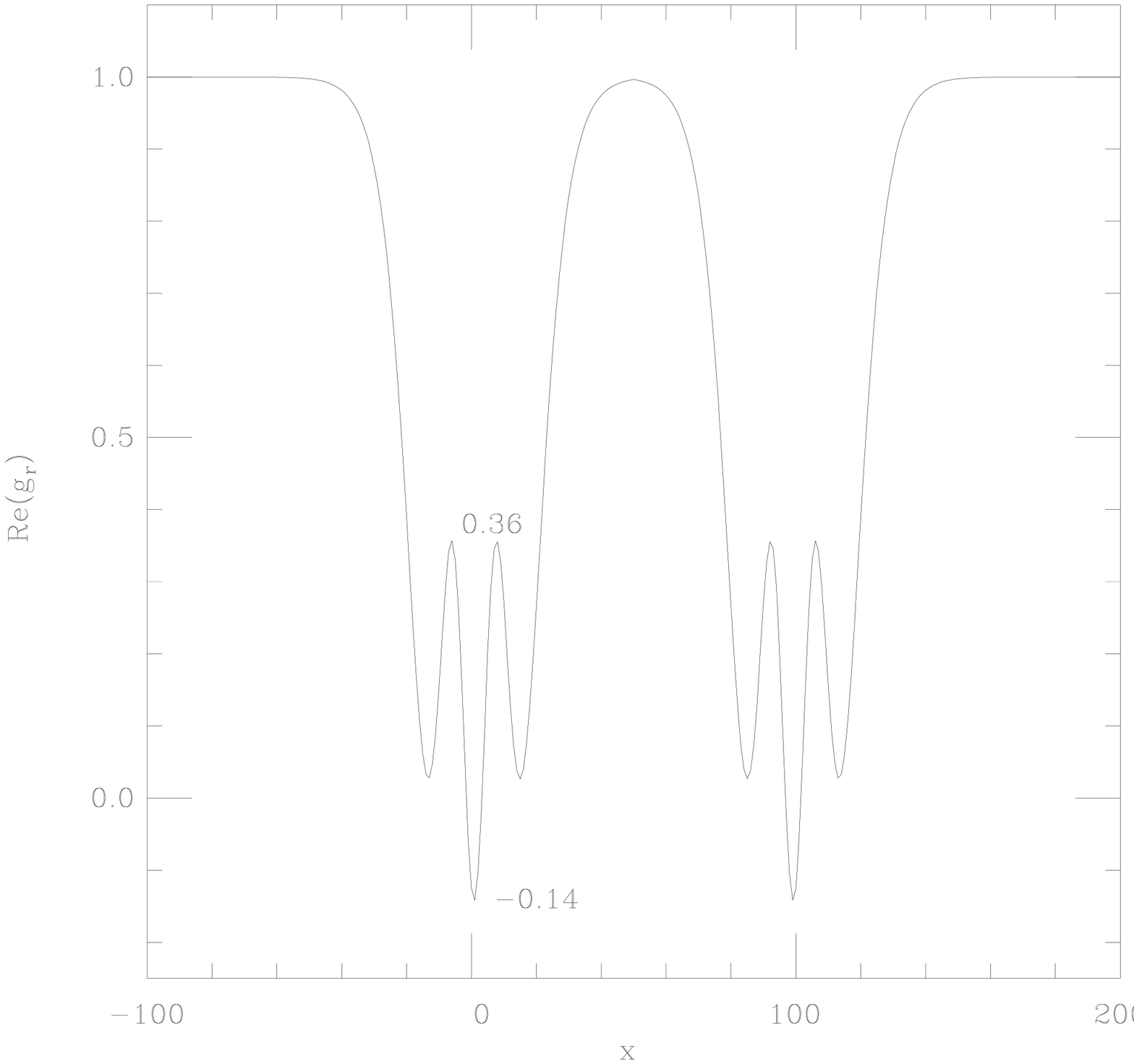}}
\vspace{-40pt}
\caption{\label{rrr}
 ${\mathrm {Re}}(g_{r})$ versus $x$ is plotted for $R = 100$ in the $14$-dimensional representation of the G($2$) gauge group. The extremums $0.36$ and -$0.14$ can be 
interpreted by the SU($3$) and SU($2$) subgroups of the G($2$) gauge group. The minimum of ${\mathrm {Re}}(g_{r})$ in the G($2$) adjoint representation corresponds to the 
nontrivial center element of the SU($2$) subgroup but the minimum of ${\mathrm {Re}}(g_{r})$ in the G($2$) fundamental representation corresponds to the nontrivial 
center element of the SU($3$) subgroup. }
\end{center}
\end{figure}
\begin{figure}
\begin{center}
\resizebox{0.87\textwidth}{!}{
\includegraphics{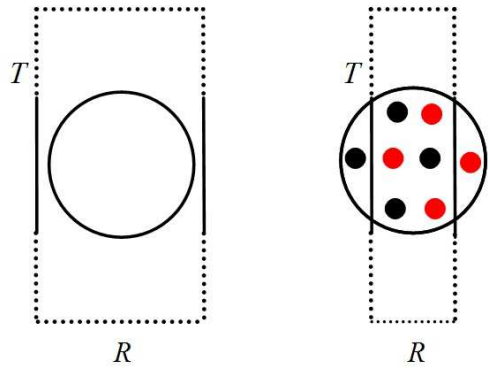}}
\caption{\label{fff}
The figure schematically shows the effect of the vacuum domain on the Wilson loop at the intermediate and large distances. At large distances where the vacuum domain is 
located completely inside the Wilson loop, it has no effect on the loop (left plot). Therefore the potential is screened. On the other hand, at 
intermediate distances where some part of the domain locates inside the loop, one gets a linear potential, probably because of the dominant role of the SU($2$) and SU($3$) 
subgroups of the G($2$) gauge group (right plot).}  
\end{center}
\end{figure}

 Next, we study the behavior of ${\mathrm {Re}}(g_{r})$ for the adjoint representation. ${\mathrm {Re}}(g_{r})$ in the adjoint representation of G($2$) can be obtained from Eqs. 
(\ref{gr1}) and (\ref{Cartan 1}) (right): 
  \begin{equation}
 {\mathrm {Re}}(g_{14})=\frac{1}{14}(4 cos(\frac{\alpha_2}{2\sqrt{24}})+2cos(\frac{\alpha_2}{\sqrt{24}})+4 cos(\frac{3\alpha_2}{2\sqrt{24}})+4)
 \end{equation}
 where the maximum flux profiles in the adjoint representation, $\alpha_1^{max}=0$ and $\alpha_2^{max}={4\pi \sqrt{24}}$, are obtained
by using the Cartan generators of Eq. (\ref{Cartan 1}) (right). Figure \ref{rrr} plots ${\mathrm {Re}}(g_{r})$ versus $x$ for $R = 100$ for 
the adjoint representation of G($2$). ${\mathrm {Re}}(g_{r})$ of the SU($2$) and SU($3$) subgroups for the adjoint representation using 
Eqs.~(\ref{Z}) and~(\ref{ZZ}) are   
 \begin{equation}
\label{min 2}
\begin{aligned}
 {min}[{\mathrm {Re}}g_{r}
(\alpha(x))]_ {SU(2)}&=\frac{1}{14}{\mathrm {Re}}(\mathrm{Tr}(e^{i\alpha \cdot
H})_\mathrm{min})=\frac{1}{14} {\mathrm {Re}}\mathrm{Tr}\left(\begin{array}{cccccc} I_{3\times3} & 0 & 0&0&0&0 \\0 &e^{i\pi}I_{4\times4} &0 &0&0&0\\ 0 & 0 & e^{i\pi}I_{4\times4}&0&0&0\\0&0&0&1&0&0\\0&0&0&0&1&0\\0&0&0&0&0&1
\end{array} \right)\\
&=\frac{1}{14}{(3 -4-4+1+1+1)}=-0.14
\end{aligned}
\end{equation}
 and
\begin{equation}
\label{min 3}
\begin{aligned}
 {min}[{\mathrm {Re}}g_{r}
(\alpha(x))]_{SU(3)}&=\frac{1}{14}{\mathrm {Re}}(\mathrm{Tr}(e^{i\alpha \cdot
H})_\mathrm{min})=\frac{1}{14}{\mathrm {Re}}\mathrm{Tr}\left(\begin{array}{ccc} e^\frac{i2\pi}{3}I_{3\times3} & 0 & 0 \\0 &e^{-\frac {i2\pi}{3}}I_{3\times3} &0 \\ 0 & 0 & I_{8\times8}
\end{array} \right)\\
&=\frac {1}{14}(-1.5-1.5 +8)=0.36.
\end{aligned}
\end{equation}
The values obtained from Eqs. (\ref{min 2}) and (\ref{min 3}) are equal with the  the extremums of ${\mathrm {Re}}(g_{r})$ of 
the G($2$) gauge group in the adjoint representation.
Therefore, the extremums -$0.14$ and $0.36$ in ${\mathrm {Re}}(g_{r})$ of the G($2$) in the adjoint representation correspond to the SU($2$) and SU($3$) subgroups.
 The absolute minimum of ${\mathrm {Re}}(g_{r})$ corresponds  to the SU($2$) subgroup, though. 

To summarize, the extremums of ${\mathrm {Re}}(g_{r})$ of the G($2$) are related to  the subgroups of G($2$). The extremums of the fundamental and adjoint representations are 
the minimums of the SU($3$) and SU($2$) subgroups. With these numerical evidences along with the fact that ${\mathrm {Re}}(g_{r})$ which comes from the vortex profile, is a factor in
Eq. (\ref{potential}) that calculates the potential between static quarks (color sources), one can conclude that the SU($2$) and SU($3$) subgroups may be responsible for the
confinement of color sources in G($2$) gauge group at intermediate distances. We recall that in the previous section, we have shown that within
a good approximation the potentials of the SU($2$) and SU($3$) subgroups are  parallel with the G($2$) potential at intermediate distances.

In the last two sections, we have discussed the possible reasons of the observed linear regime of the G($2$) gauge group. Figure \ref{fff} schematically plots the 
possible behavior of the vacuum domain at large and intermediate distances.

\section{conclusion}
 
According to the center vortex model, the nontrivial center elements are responsible for the confinement. But numerical lattice calculations show a 
linear regime for G($2$) gauge theory which does not have any nontrivial center element. On the other hand, we have observed a linear regime for    
 G($2$) gauge theory using a domain model, which modifies the thick center vortex model by adding a contribution for the trivial center element in 
addition to the nontrivial center elements. In this article, we investigate the possible reasons for the confinement in the G($2$) gauge group. In the 
confinement regime of G($2$), we observe two linear regimes where the first one agrees qualitatively with the Casimir scaling. 

To interpret the second linear regime, the potentials of SU($2$) and SU($3$) subgroups are compared with the G($2$) potential in the 
confinement regime. The 
second linear regime of the G($2$) gauge theory is roughly  parallel 
with the linear regimes of SU($2$) and SU($3$) subgroups. Then, we study ${\mathrm {Re}}(g_{r})$, related to the vortex profile, for G($2$) and its subgroups. 
 In the SU($N$) gauge group, ${\mathrm {Re}}(g_{r})$ changes between $1$ and some values corresponding to the nontrivial center elements. 
But in the G($2$) gauge group, ${\mathrm {Re}}(g_{r})$ changes between $1$ and  some values which are explained  in terms of the
G($2$) subgroup center elements. We have learned that  these values are equal with the ${min}[{\mathrm {Re}g_{r}}]$ 
of the SU($2$) and SU($3$) subgroups. We have argued that SU($2$) and SU($3$) subgroups of G($2$) have important roles in observing confinement in G($2$) gauge group.

\section{\boldmath Acknowledgments}

 We are grateful to the research council of the University of Tehran for
supporting this study.

\appendix*

\section{ DECOMPOSITION OF G($2$) REPRESENTATIONS TO ITS SU($2$) AND THE WEIGHT DIAGRAMS} 

In this Appendix, we obtain the decomposition of $7$ and $14$ representations of the G($2$) gauge group into their SU($2$) subgroups with the help of the generators ${H}_{8}$, ${H}_{9}$, and ${H}_{10}$ group of Eq. (\ref{gen}).
${H}_{8}$ is obtained from Eq. (\ref{Cartan 1}): 
\begin{equation}
\label{Cartan A}
H_8= \frac{1}{2\sqrt{6}} \left( \begin{array}{ccccccc} \ 1 &1 &-2 &-1 &-1 & 2&0\end{array} \right) \ = \frac{1}{\sqrt{6}} \left( \begin{array}{ccccccc}\frac{1}{2} &\frac{1}{2} &-1 &-\frac{1}{2} &-\frac{1}{2} & 1&0 \
\end{array} \right),
\end{equation}
where only the elements of the diagonal are reported. 
Diagonal elements of the Cartan generators of SU($2$) of representations $j=1/2,1,3/2,...$ are
\begin{equation}
\label{Cartan B}
 \sigma_3^{(j=\frac{1}{2})}=\left( \begin{array}{cc} \ \frac{1}{2} &-\ \frac{1}{2} \end{array} \right),
  \sigma_3^{(j=1)}=\left( \begin{array}{ccc} \ 1 &0&-1 \end{array} \right), 
   \sigma_3^{(j=\frac{3}{2})}=\left( \begin{array}{cccc} \ \frac{3}{2}&\ \frac{1}{2} &-\ \frac {1}{2}&-\ \frac{3}{2} \end{array} \right),... \ .
\end{equation}
\begin{figure}
\begin{center}
\resizebox{0.6\textwidth}{!}{
\includegraphics{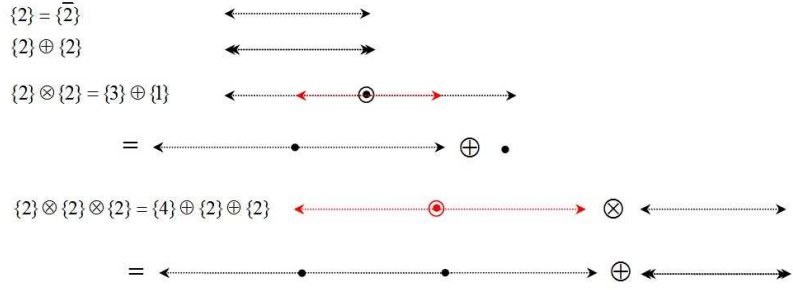}}
\caption{\label{m}The weight diagrams of some SU($2$) representations are constructed. The circles around a point signify the degeneracy of the states. The weight diagrams of all of the SU($2$) representations are one-dimensional because the dimension of the SU($2$) fundamental representation is $1$. We illustrate the graphical construction of 
$2 \otimes 2$, for example. First we place a $2$ $(\leftrightarrow)$ at the origin. Then, we put two $2$'s  $(\leftrightarrow)$ on the tips at the 
original $2$ such that the center of the $2$'s  sits on the tips of the original $2$. The top plot shows the resulted representations: $\{3\} \oplus \{1\}$.
 Similar methods can be used for other representations.}
\end{center}
\end{figure}
\begin{figure}
\begin{center}
\resizebox{0.87\textwidth}{!}{
\includegraphics{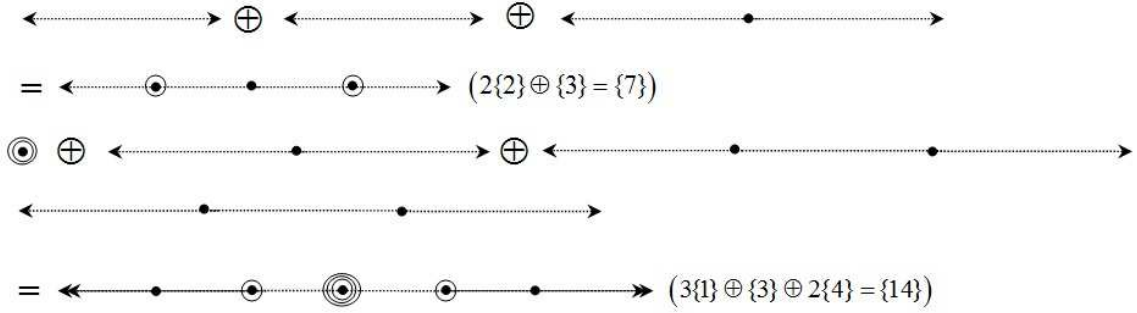} }
\caption{\label{m12}The weight diagrams of seven- and $14$-dimensional representations for the SU($2$) subgroup of G($2$). }
\end{center}
\end{figure}

Comparing these Cartans with the Cartan generator ${H}_{8}$, one observes that ${H}_{8}$ is constructed from two $\sigma_3^{(j=\frac{1}{2})}$ and one $\sigma_3^{(j=1)}$. Therefore, 
it is confirmed that the fundamental representation of G($2$) can be decomposed to its SU($2$) subgroups as the following:
\begin{equation}
\label{7t}
\{7\} = 2\{2\} \oplus \{3\}.
\end{equation}
For the adjoint representation, ${H}_{8}$ is obtained from Eq. (\ref{Cartan 1}): 
\begin{equation}
\label{Cartan C}
\begin{aligned}
H_8&= \frac{1}{\sqrt{8}} \left( \begin{array}{cccccccccccccc} \ \frac{1}{2\sqrt{3}} &\ \frac{1}{2\sqrt{3}} &-\ \frac{1}{\sqrt{3}} &-\ \frac{1}{2\sqrt{3}} &-\ \frac{1}{2\sqrt{3}} & \ \frac{1}{\sqrt{3}}&0&0&\ \frac{3}{2\sqrt{3}}&0&0&\ \frac{3}{2\sqrt{3}}&-\ \frac{3}{2\sqrt{3}}&-\ \frac{3}{2\sqrt{3}}\end{array} \right)\\
&= \frac{1}{\sqrt{24}} \left( \begin{array}{cccccccccccccc} \ \frac{1}{2} &\ \frac{1}{2} &-1 &-\ \frac{1}{2} &-\ \frac{1}{2} & 1&0&0&\ \frac{3}{2}&0&0&\ \frac{3}{2}&-\ \frac{3}{2}&-\ \frac{3}{2}\ 
\end{array} \right).
\end{aligned}
\end{equation}
Comparing Eq. (\ref{Cartan C}) with the generator of SU($2$) of representations $j=1,1/2,3/2,...$, it is clear that ${H}_{8}$ is constructed from two $\sigma_3^{(j=\frac{3}{2})}$, one $\sigma_3^{(j=1)}$, and three $0$. Therefore decomposition of the G($2$) adjoint representation into its SU($2$) subgroup is given by
\begin{equation}
\label{Cartan j}
\{14\} = 3\{1\} \oplus \{3\} \oplus 2\{4\}.
\end{equation}
To obtain the weight diagram of $7$ and $14$ representations of the SU($2$) subgroup of G($2$), we construct weight diagrams of many SU($2$) representations shown in 
Fig. \ref{m}. The weight diagrams of the fundamental and the adjoint representations of the SU($2$) subgroup of G($2$) are also shown in Fig. \ref{m12}.

\end{document}